\documentclass[conference]{IEEEtran}
\IEEEoverridecommandlockouts
\IEEEpubid{\makebox[\columnwidth]{\hfill}}

\usepackage{cite}
\usepackage{amsmath,amssymb,amsfonts}
\usepackage{graphicx}
\usepackage{textcomp}
\usepackage{algorithm}
\usepackage{algpseudocode}
\usepackage[dvipsnames,svgnames]{xcolor}
\PassOptionsToPackage{hyphens}{url}
\usepackage{hyperref}
\usepackage{stfloats}
\usepackage{tikz}
\usetikzlibrary{positioning,calc}
\usepackage{placeins}
\makeatletter
\renewcommand{\fnum@algorithm}{\footnotesize Algorithm~\thealgorithm}
\makeatother
\algrenewcommand\algorithmiccomment[1]{\hfill\(\triangleright\) #1}
\algrenewcommand\alglinenumber[1]{\scriptsize #1}

\def\BibTeX{{\rm B\kern-.05em{\sc i\kern-.025em b}\kern-.08em
    T\kern-.1667em\lower.7ex\hbox{E}\kern-.125emX}}
\begin{document}

\title{\color{black}{Framework Implementation Maturity in Blockchain-Based Third-Party Compliance Assessment} \\
}

\author{
\IEEEauthorblockN{
Jemima Owusu-Tweneboah\IEEEauthorrefmark{1},
Amani Altarawneh\IEEEauthorrefmark{1},
Deepti Gupta\IEEEauthorrefmark{2},
Maria Luisa Figueroa\IEEEauthorrefmark{2}}
\IEEEauthorblockA{\IEEEauthorrefmark{1}\textit{Computer Science Department} \\
\textit{Tennessee Technological University}, United States \\
\{jowusutwe42, aaltarwaneh\}@tntech.edu}
\IEEEauthorblockA{\IEEEauthorrefmark{2}\textit{Computer Information Systems Department} \\
\textit{Texas A\&M University Central}, United States \\
\{d.gupta, maria.figueroa\}@tamuct.edu}
}
 
\maketitle

%
%
%
\begin{abstract}
Cybersecurity and privacy frameworks such as NIST SP~800--53, ISO/IEC~27001, GDPR, and HIPAA are widely used to guide organizational security posture and regulatory compliance. In practice, however, framework adoption is often assessed through point-in-time audits, self-attestations, and fragmented evidence reviews, providing limited assurance that controls are consistently implemented, independently validated, and sustained over time, particularly in environments that rely on third-party vendors. These limitations are amplified in multi-vendor ecosystems, such as healthcare remote patient monitoring (RPM), where compliance obligations span organizational boundaries and assessments are conducted by multiple independent assessors. This paper investigates how permissioned blockchain systems can support framework implementation maturity measurement rather than static compliance verification. We propose a blockchain-based Third-Party Risk Assessment (TPRA) framework that operationalizes assessment workflows, enforces multi-party governance, and preserves longitudinal assessment state using programmable smart contracts. Building on this framework, we introduce a set of evaluation metrics and a qualitative maturity model designed to assess whether compliance controls are verifiably implemented, governed, and sustained across repeated assessment cycles.
Using healthcare Remote Patient Monitoring (RPM) as a representative but non-exhaustive case study, we conduct a scenario-driven, demonstrative evaluation to illustrate how maturity metrics can be applied in practice. The evaluation highlights how immutable assessment records, cryptographic evidence anchoring, and governed assessor consensus enable reproducible, auditable, and longitudinal compliance reasoning that is not supported by traditional audit or centralized Governance, Risk, and Compliance (GRC) approaches. Using scenario-driven evaluation, we illustrate how TPRA supports a governed, evidence-grounded, and maturity-oriented approach to third-party compliance assessment in regulated multi-organization settings.

\end{abstract}


\section{Introduction}
\textcolor{black}{Third-party vendors play a critical role in the operation of regulated systems~\cite{gan2024large,ilori2022cybersecurity}, particularly in healthcare and other critical information infrastructure (CII) domains. Services such as cloud-hosted data storage (e.g., AWS/Azure healthcare workloads), analytics platforms, and Internet-of-Health (IoH) device ecosystems (e.g., connected glucose monitors and cardiac telemetry systems) are routinely operated by external providers.}

\textcolor{black}{Consider a connected health device that transmits regulated patient telemetry to a cloud backend for storage and analytics. Although vendors may claim HIPAA/GDPR alignment during system design, implementation and iterative updates can introduce gaps across third-party components (e.g., misconfigured storage permissions, incomplete audit logging, weak access governance, or inconsistent key-management practices). Without end-to-end governance and independent verification across stakeholders, compliance degrades into a point-in-time artifact rather than an enforceable, auditable system property.}

\textcolor{black}{Although cybersecurity frameworks such as NIST SP~800-53 define extensive control requirements~\cite{ni2025recent,andrew2023blockchain,syafrizal2020analysis,wang2024survey}, vendor compliance is still evaluated largely through point-in-time audits, self-attestations, and document-based assessments~\cite{hassani2024rethinking, ilori2022cybersecurity, sabillon2022cybersecurity}. Most programs remain question-driven (e.g., questionnaires and attestations), and recent work has explored blockchain to improve auditability, evidence integrity, and accountability~\cite{andrew2023blockchain,ni2025recent}. However, existing approaches largely record artifacts or verify isolated attestations~\cite{islam2025logstamping}, offering limited support for \emph{measuring framework implementation maturity} across repeated assessment cycles in multi-assessor, governed settings.}

In this paper, \emph{maturity} is the degree to which control enforcement is consistently evidenced and reproducible across assessment cycles from system execution, rather than narrative reports. Because current assessments are largely point-in-time and report-centric, outcomes are hard to reproduce, compare, or validate longitudinally (from one assessment period to the next over time). We address this gap by reframing third-party compliance assessment as a \emph{maturity-oriented evaluation problem}, supported by a permissioned-ledger protocol that preserves assessment state across cycles.

TPRA goes beyond logging compliance artifacts by introducing a \emph{ledger-enforced assessment state machine}. Evidence is cryptographically bound to an immutable assessment context, independent assessor judgments are recorded and aggregated via a deterministic rule, and outcome finalization is governed by endorsement policies that specify authorized decision-makers.

This paper makes the following contributions toward maturity-oriented third-party compliance assessment:
\begin{itemize}
    \item \textcolor{black}{We formalize a repeatable maturity-oriented evaluation model for measuring framework implementation across assessment cycles.}
    \item \textcolor{black}{We design a ledger-enforced smart-contract protocol that executes assessment actions and preserves state across cycles.}
    \item \textcolor{black}{We define a set of evidence-grounded maturity metrics, derived from protocol artifacts, for evaluating governed third-party compliance assessments over time.}
    \item \textcolor{black}{We present a multi-organization threat model and provide proof sketches for integrity binding, non-repudiation, and endorsement-gated finalization.}
    \item \textcolor{black}{We evaluate the approach in a healthcare RPM case study and discuss generalization to other regulated, multi-vendor domains.}
\end{itemize}

\textcolor{black}{We prioritize maturity-oriented evaluation and governance enforcement over domain-specific optimization. We validate feasibility through end-to-end workflow execution, system-enforced evidence handling, and maturity metrics derived from protocol traces; large-scale deployment benchmarking remains future work.}

\section{Related Work}
Related work spans (i) third-party assessment and maturity practice, (ii) non-blockchain compliance automation and continuous compliance, and (iii) blockchain-based mechanisms for auditability and governance. Across these areas, prior work improves standardization, efficiency, or tamper-evidence, but offers limited support for \emph{maturity-oriented} evaluation that is reproducible across assessors and sustained across repeated cycles. Table~\ref{tab:rw-comparison} summarizes representative approaches and contrasts their support for tamper-evident auditability, governed multi-assessor finalization, reproducible outcomes, longitudinal state, and evidence-grounded metrics.

\begin{table*}[!t]
\centering
\caption{Qualitative comparison of related approaches and TPRA across five dimensions.}
\label{tab:rw-comparison}
\renewcommand{\arraystretch}{1.05}
\setlength{\tabcolsep}{5pt}
\begin{tabular}{|p{2.8cm}|p{2.6cm}|p{3.6cm}|c|c|c|c|c|}
\hline
\textbf{Work} &
\textbf{Domain} &
\textbf{Core Technology / Procedure} &
\textbf{TE Audit} &
\textbf{MA Gov} &
\textbf{Reprod} &
\textbf{Long. State} &
\textbf{EG Metrics} \\
\hline

Ilori et al.~(2022) \cite{ilori2022cybersecurity} &
Cybersecurity auditing &
Manual audits, regulatory processes &
-- & -- & -- & -- & -- \\
\hline

Aljumaiah et al.~(2025) \cite{aljumaiah2025analyzing} &
Critical infrastructure &
NIST CSF mapping, qualitative risk analysis &
-- & -- & -- & -- & -- \\
\hline

Angermeir et al.~(2024) \cite{angermeir2024continuous} &
Continuous compliance &
Automation roadmap, control mappings, verification pipelines &
-- & -- & \textcircled{--} & -- & \textcircled{--} \\
\hline

Silva et al.~(2024) \cite{silva2024automation} &
Cloud/IaaS vulnerability management, continuous compliance &
Automated control operationalization + tool execution; CVSS-based metrics &
-- & -- & \checkmark & -- & \checkmark \\
\hline

Avancha et al.~(2024) \cite{avancha2024blockchain} &
IT vendor management &
Blockchain + smart contracts (survey/positioning) &
\textcircled{--} & -- & -- & -- & -- \\
\hline

Gupta et al.~(2024) \cite{gupta2024blockchain} &
Third-party vendor risk &
Blockchain + smart contracts; continuous monitoring of controls (case study) &
\textcircled{--} & -- & \textcircled{--} & -- & -- \\
\hline

Adlam et al.~(2020) \cite{adlam2020permissioned} &
Healthcare EHR auditing &
Hyperledger Fabric audit-log chaincode/prototype &
\checkmark & -- & -- & -- & -- \\
\hline


Psarra et al.~(2024) \cite{psarra2024permissioned} &
Healthcare access control &
Permissioned Fabric + ML (LSTM) + fuzzy logic; latency eval &
\checkmark & -- & -- & \textcircled{--} & -- \\
\hline

Shu et al.~(2021) \cite{shu2021blockchain} &
Cloud storage auditing &
Blockchain-based decentralized public auditing; DAO/SC mechanism &
\checkmark & -- & \textcircled{--} & -- & -- \\
\hline

Banerjee et al.~(2025) \cite{banerjee2025cumulus} &
Cloud data audit (privacy) &
Smart contracts + state channels; UC security proof; Ethereum prototype &
\checkmark & -- & \checkmark & -- & \textcircled{--} \\
\hline

Islam et al.~(2025) \cite{islam2025logstamping} &
Log auditing (large-scale systems) &
Hybrid on-/off-chain + IPFS + smart-contract validation; perf eval &
\checkmark & -- & -- & -- & -- \\
\hline



Kong et al.~(2025) \cite{kong2025research} &
Enterprise internal audit &
Blockchain-based audit process + smart contracts (automation framing) &
\textcircled{--} & -- & -- & -- & -- \\
\hline

Chung et al.~(2022) \cite{chung2022applicability} &
Healthcare project risk mgmt &
Blockchain-based collaborative risk info model + initial prototype &
\textcircled{--} & -- & \textcircled{--} & -- & -- \\
\hline

Anderson~(2018) \cite{anderson2018securing} &
Healthcare EHR audit logs &
Hyperledger Fabric + chaincode to standardize/consolidate audit logs (prototype) &
\checkmark & -- & \checkmark & -- & -- \\
\hline

Azaria et al.~(2016) \cite{azaria2016medrec} &
Healthcare permissioning &
Blockchain-based permission management + immutable access log (prototype) &
\checkmark & -- & \checkmark & -- & -- \\
\hline

Barbaria et al.~(2025) \cite{barbaria2025advancing} &
Healthcare data exchange compliance &
Fabric + IPFS + FHIR; smart contracts for consent/policy enforcement (architecture) &
\checkmark & -- & \checkmark & -- & -- \\
\hline

Ullah et al.~(2024) \cite{ullah2024blockchain} &
Healthcare EHR auditing &
Smart contracts + PBAC policy verification; immutable audit trail/logs (implementation) &
\checkmark & -- & \checkmark & -- & -- \\
\hline

Chatziamanetoglou et al.~(2023) \cite{chatziamanetoglou2023blockchain} &
Security configuration management &
Permissioned blockchain + smart-contract RBAC + lifecycle provenance &
\checkmark & -- & \checkmark & -- & -- \\
\hline

Androulaki et al.~(2019) \cite{androulaki2019endorsement} &
Fabric governance primitives &
Endorsement model + state-based endorsement + security analysis/benchmarks &
\checkmark & \checkmark & \checkmark & -- & -- \\
\hline

\textbf{TPRA (This work)} &
Third-party compliance &
Permissioned blockchain as assessment protocol &
\checkmark & \checkmark & \checkmark & \checkmark & \checkmark \\
\hline

\end{tabular}

\vspace{2pt}
\footnotesize
\textbf{Legend:}
TE Audit = tamper-evident cross-organization auditability;
MA Gov = multi-assessor governed finalization;
Reprod = reproducible outcome construction;
Long. State = longitudinal assessment state;
EG Metrics = evidence-grounded metrics.
\checkmark\ = Full support;
\textcircled{--} = Partial or implicit support;
-- = Not supported.

\end{table*}

\paragraph{Third-Party Assessment and Maturity Practice}
Third-party cyber risk assessment is commonly treated as an organizational process shaped by context and assessor judgment. Slapni{\v{c}}ar et al.\ develop a process theory of supplier cyber risk assessment using qualitative analysis and expert surveys, emphasizing practice variability even under standardized frameworks and the resulting challenges for reproducibility across time and assessors~\cite{slapnicar2025process}. Formal maturity programs such as the Cybersecurity Maturity Model Certification (CMMC) define maturity levels and assessment objectives~\cite{strohmier2022cybersecurity}, but outcomes are typically captured as reports or attestations with limited cryptographic binding between evidence, assessor judgments, and reassessments. Enterprise audit modernization motivates integrity and automation: Kong proposes a blockchain-based internal audit process in which smart contracts automate audit procedures and immutability/timestamping strengthen evidence credibility~\cite{kong2025research}, and Avancha et al.\ survey blockchain-based vendor management to motivate transparency, audit trails, and automation of contract/credential verification~\cite{avancha2024blockchain}. These works motivate ledger-backed provenance, but do not specify an executable third-party assessment protocol with preserved independent assessor inputs and governed finalization across recurring cycles, which TPRA operationalizes.

\paragraph{Non-Blockchain Compliance Automation and Continuous Compliance}
Non-blockchain approaches reduce cost and subjectivity via automation and continuous compliance. Angermeir et al.\ define continuous security compliance, summarize challenges through a tertiary study, and propose a roadmap based on control mappings and verification pipelines~\cite{angermeir2024continuous}. Silva et al.\ operationalize controls for cloud vulnerability management by decomposing CIS/NIST-style recommendations into actionable conditions, tying checks to automated tool execution, and reporting compliance using CVSS-derived measures (e.g., node verification)~\cite{silva2024automation}. These approaches are effective within single administrative boundaries or centralized platforms, where a unified operator can run tools and consolidate outcomes; in third-party settings, standardized questionnaires and templates remain document-centric and typically rely on trusted intermediaries, limiting tamper-evident provenance and governed multi-party outcome construction. Chung and Caldas highlight similar multi-stakeholder coordination challenges and propose a conceptual blockchain-based risk-information model with an initial prototype, but focus on risk-event collaboration rather than protocol-enforced assessment semantics and longitudinal maturity measurement~\cite{chung2022applicability}.

\paragraph{Blockchain-Based Auditability and Governance}
Permissioned blockchains are frequently used to strengthen integrity, accountability, and access control in regulated workflows. Hyperledger Fabric provides authenticated membership, endorsement policies, and a tamper-evident ledger, enabling governance-enforced transaction validity~\cite{androulaki2018hyperledger}. Representative systems anchor logs or audit artifacts on-chain: LogStamping proposes blockchain-based log auditing with hybrid on-chain/off-chain storage, InterPlanetary File System (IPFS) for scalability, and smart-contract validation~\cite{islam2025logstamping}, while Anderson’s \textit{AuditChain} uses Fabric chaincode to standardize audit-log structure and consolidate audit data across healthcare organizations~\cite{anderson2018securing}. Healthcare-oriented systems similarly apply smart contracts to access governance, consent enforcement, and immutable audit logging, including architectures built on Hyperledger Fabric, IPFS (InterPlanetary File System), and FHIR (Fast Healthcare Interoperability Resources), as well as PBAC (Policy-Based Access Control)-based auditing and related provenance mechanisms~\cite{tawfik2025achealthchain,barbaria2025advancing,ullah2024blockchain,gupta2024blockchain,chatziamanetoglou2023blockchain,azaria2016medrec,adlam2020permissioned,shuaib2022secure,psarra2024permissioned}. Cloud-storage auditing targets integrity verification and dispute handling: Shu et al.\ propose decentralized public auditing and Banerjee et al.\ (Cumulus) use smart contracts as a dispute mediator for privacy-preserving audits with reduced on-chain overhead via state channels~\cite{shu2021blockchain,banerjee2025cumulus}. Finally, platform-level studies formalize why Fabric supports governed multi-party workflows and motivate endorsement-gated validity, including analyses of accountability, endorsement models, and policy performance~\cite{graf2020accountability,androulaki2019endorsement,piao2023performance}. However, most prior work treats the ledger as a logging, access-control, or integrity-verification substrate rather than an assessment protocol that encodes lifecycle semantics, preserves independent assessor judgments, and enforces governed outcome construction across repeated cycles.

\paragraph{Research Gap and Positioning}
Across procedural maturity programs, non-blockchain compliance automation, and blockchain-based audit systems, a common gap remains: prior work emphasizes verification, artifact integrity, or workflow efficiency, but does not provide a unified mechanism for \emph{measuring framework implementation maturity} as a reproducible, multi-assessor, longitudinal process. Integrated cybersecurity auditing models argue for unifying compliance, risk management, and governance~\cite{ilori2022cybersecurity,sabillon2022cybersecurity}, but remain largely procedural and do not specify enforceable assessment semantics or protocol-level governance, as reflected in Table~\ref{tab:rw-comparison}. There, a dimension is coded as \emph{Full} when it is explicitly supported as a primary feature of the architecture, protocol, or evaluation; \emph{Partial} when it is present only in limited or indirect form; and \emph{None} when it is absent or not substantively developed. In contrast, TPRA positions a permissioned blockchain as an \emph{enforcement substrate} for assessment protocols by binding assessor participation, evidence anchoring, endorsement-gated finalization, and longitudinal state transitions into an auditable history suitable for maturity-oriented evaluation.

\section{System Context}
\label{sec:context}
Remote Patient Monitoring (RPM) systems enable continuous collection and analysis of patient physiological data using networked medical devices and cloud-based platforms. Typical deployments integrate IoH sensing devices (e.g., blood pressure monitors, glucose sensors, cardiac monitors), vendor-operated cloud infrastructures for data ingestion and analytics, and clinical dashboards used by healthcare providers for monitoring and decision support. In practice, RPM device data is transmitted to third-party cloud services for storage, preprocessing, and analytics before clinician access, and multiple external vendors (device manufacturers, cloud providers, analytics platforms, and integrators) participate in processing sensitive health information and in implementing security controls~\cite{abouelmehdi2018big, fernandez2013security, reisman2017ehrs}. This creates cross-organization dependencies in which critical controls may be operated outside the healthcare provider’s administrative boundary.

\paragraph{Third-Party Accountability and Compliance Asymmetry}Despite this distribution of operational responsibility, healthcare providers remain accountable for compliance with regulatory frameworks such as HIPAA and control baselines such as NIST SP~800-53. However, these frameworks specify \emph{what} controls should exist more than \emph{how} their ongoing enforcement should be verified across independently operated vendor environments. Consequently, third-party assessment practice remains largely questionnaire- and document-driven, offering limited verifiable visibility into whether controls are sustained over time versus satisfied at assessment time.

\paragraph{Assessment and Governance Challenges}RPM highlights governance challenges in third-party assessment. Vendors are often evaluated by multiple independent assessors with different scopes or technical specializations, and assessment outcomes may vary due to expert judgment and evidence interpretation even under the same framework~\cite{slapnicar2025process, keskin2021cyber}. Further, assessments are typically periodic, producing point-in-time snapshots; changes in vendor configurations, operational practices, and threat conditions between cycles are not systematically captured, limiting reproducibility and longitudinal comparability.

\paragraph{Why Healthcare RPM Serves as a Stress Test}We use healthcare RPM as a stringent but representative stress test for maturity-oriented third-party assessment because it is regulated, multi-vendor, and operationally dynamic. While our evaluation is instantiated in RPM, the assessment constructs generalize to other regulated multi-vendor settings (e.g., financial services, critical infrastructure, and cloud enterprise systems) where independent validation, governed outcome construction, and cross-cycle comparability are required.

\FloatBarrier

\section{Third-Party Risk Assessment Framework Architecture and Design}
\label{sec:arch}
\textcolor{black}{This section presents TPRA as a permissioned, multi-organization blockchain application implemented on Hyperledger Fabric. Fabric supports execute--order--validate transaction processing, endorsement-policy governance, and channel-based data partitioning across organizations~\cite{androulaki2018hyperledger}. These primitives match third-party assessment settings where (i) roles are institutionally known (regulator/assessor/vendor), (ii) state changes require multi-party authorization, and (iii) the system must preserve an immutable, reconstructable assessment history across cycles.}

\textcolor{black}{TPRA operationalizes third-party compliance assessment as a \textit{governed protocol} executed on a permissioned blockchain, as shown in \textcolor{black}{Figure \ref{fig:tpra_architecture}}. Assessment actions (creation, evidence anchoring, voting, and finalization) are encoded as chaincode transactions whose effects are constrained by endorsement policies and role-based authorization. We refer to the TPRA workflow (i.e., the chaincode-enforced assessment lifecycle and endorsement-gated governance) as the \emph{TPRA protocol}}.

\renewcommand{\thefigure}{\arabic{figure}}
\setcounter{figure}{0}
\begin{figure*}[!t]
\centering
\makebox[\textwidth][c]{%
\includegraphics[width=0.9\textwidth]{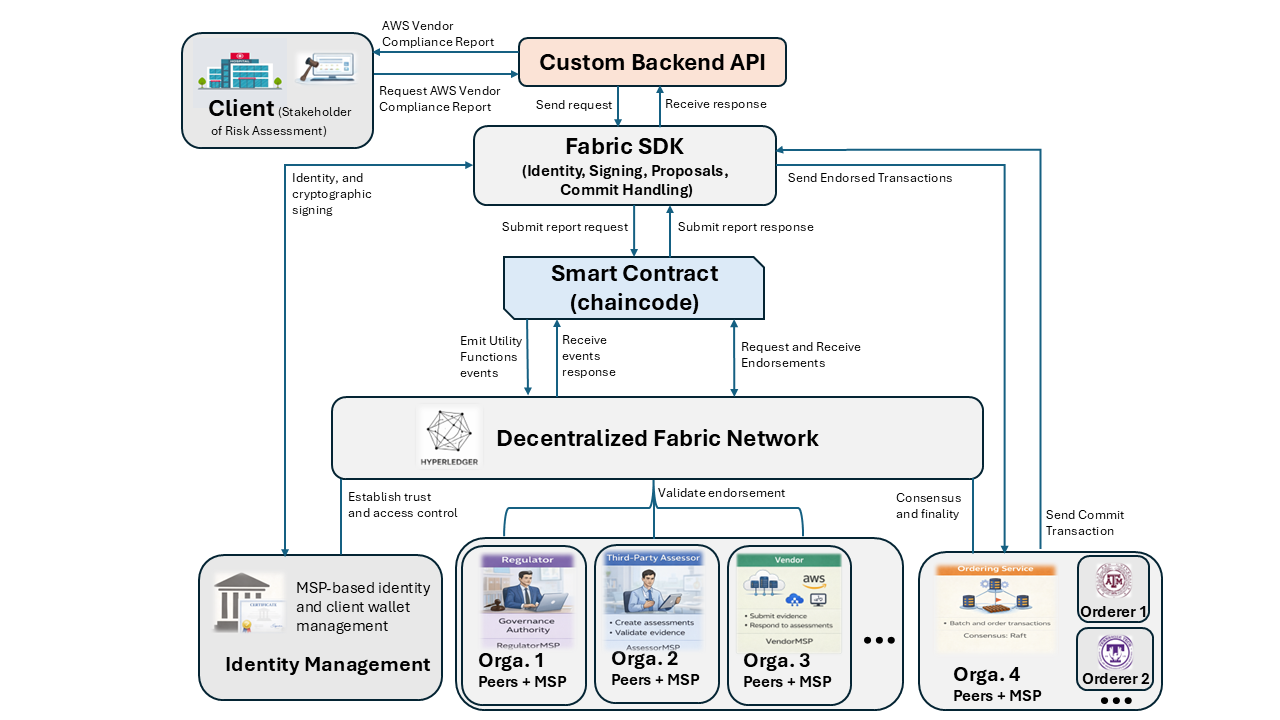}}
\caption{System overview of TPRA on Hyperledger Fabric. A client invokes assessment functions via a backend API and Fabric SDK using MSP-authenticated identities. Chaincode transactions implement the assessment lifecycle (e.g., vendor registration, framework/question initialization, assessment creation, evidence anchoring, assessor voting, governed finalization, etc.). Transactions are endorsed by designated organizations per policy, ordered by an ordering service, validated, and committed, producing an immutable assessment state and verifiable audit artifact returned to the client.}
\label{fig:tpra_architecture}
\end{figure*}

\subsection{System Model and Actors}
\label{sec:arch:actors}
\textcolor{black}{We consider a regulated third-party ecosystem in which compliance outcomes depend on controls implemented within vendor environments. TPRA models three primary organizations (and logical roles): the \emph{\textbf{Regulator}} (governance authority), which publishes assessment directives and canonical control statements/questions (e.g., HIPAA-aligned requirements mapped to NIST families) and participates in governance decisions that finalize or certify outcomes; \emph{\textbf{Third-Party Assessor(s)}}, which act as independent evaluators that validate evidence, score control implementation, and contribute votes used to construct outcomes; and \emph{\textbf{Vendor(s)}}, which are third-party entities (e.g., device manufacturer, cloud provider) that submit compliance responses and register evidence references. TPRA targets environments with multiple assessors (often specialized by control domain), where reconciliation of conflicting judgments and preservation of longitudinal assessment state are first-class requirements~\cite{slapnicar2025process}.}


\subsection{Fabric Network, Channel Governance, and MSP-Based Role Enforcement}
\label{sec:arch:fabric}
\textcolor{black}{TPRA is deployed as a Fabric network composed of multiple \emph{organizations} operating endorsing \emph{peers} and a crash-fault-tolerant \emph{ordering service}. A Fabric \emph{channel} defines the consortium ledger namespace, membership (MSP roots), and application policies~\cite{androulaki2018hyperledger}. In TPRA, RegulatorOrg and AssessorOrg run endorsing peers to enforce governance separation, while VendorOrgs may run peers for proposal submission and ledger verification.}
\textcolor{black}{Fabric’s Membership Service Provider (MSP) binds identities to organizations via X.509 certificates. TPRA uses MSP-backed identities to (i) authenticate invokers, (ii) attribute actions to organizations, and (iii) enforce role-based access control at the chaincode layer. Each organization issues identities via its Certificate Authority (CA); clients sign proposals and endorsers sign responses, enabling non-repudiable attribution of who invoked an action and which organizations approved it. Application roles (Regulator/Assessor/Vendor) are derived deterministically from MSP ID and certificate attributes, and chaincode authorizes invocation via \texttt{GetMSPID()} and invoker identity checks (e.g., vendors cannot finalize outcomes; assessors cannot publish regulator directives)~\cite{androulaki2018hyperledger}.}
\textcolor{black}{The ordering service establishes a total order of transactions per channel and disseminates blocks to peers. Stronger adversaries and accountability considerations are handled in the threat model (Section~\ref{sec:threat}).}

\subsection{Ledger Data Model and Chaincode Workflow}
\label{sec:arch:chaincode}
\textcolor{black}{TPRA is specified as a set of Fabric chaincode transactions that operationalize the assessment lifecycle. During endorsement, proposals execute against a peer’s current world state to produce a read-set/write-set; only validated write-sets are committed. TPRA stores structured world-state objects such as: \textbf{FrameworkDirective} (canonical statements/questions, tags, version metadata), \textbf{Assessment} (vendor ID, scope, framework version, status, deadlines), \textbf{EvidenceRecord} (hash, pointer/URI, timestamps, control mapping), \textbf{AssessorVote} (per-assessor score and rationale metadata without sensitive evidence), and \textbf{AssessmentOutcome} (aggregated results and audit pointers).}

The contract exposes transactions corresponding to key lifecycle stages (representative interface points: registerVendor, createAssessment, registerEvidence, submitAssessment, submitAssessorVote, and completeAssessmentReview ~\cite{gupta2024blockchain}). To support reproducibility under Fabric endorsement, TPRA chaincode is deterministic: state transitions and aggregation rules avoid non-deterministic sources and encode policy-relevant parameters in on-chain state~\cite{androulaki2018hyperledger}.

\subsection{Transaction Processing and Endorsement as Enforceable Governance}
\label{sec:arch:txflow}
Fabric separates transaction processing into endorsement (simulation), ordering, and validation/commit. A client sends a proposal to the required endorsing peers; each endorser authenticates the invoker, executes chaincode, returns a read-set/write-set, and signs the response. After collecting sufficient endorsements, the client submits the transaction to the ordering service for total ordering and block formation. Peers then validate each transaction by (i) checking endorsement-policy satisfaction, (ii) verifying endorsement signatures, and (iii) performing Multi-Version Concurrency Control (MVCC) validation against read-set versions; valid write-sets update world state while invalid transactions remain recorded but do not update state.

\textcolor{black}{TPRA encodes governance by requiring multi-organization endorsements for state-changing actions, making authorization \emph{enforceable} rather than procedural. Examples include: Regulator+Assessor endorsements for publishing directives or finalizing outcomes; assessor quorum for recording votes; and (optionally) Vendor+Assessor co-endorsement for evidence-registration flows requiring co-acknowledgment. \emph{Transaction-level workflow} observations are discussed in Section \ref{sec:eval}, and endorsement-based security guarantees are formalized in Section~\ref{sec:threat}.}

\subsection{Evidence Handling and Privacy-Preserving Storage}
\label{sec:arch:evidence}
\textcolor{black}{TPRA separates \emph{evidence content} from \emph{evidence accountability}. Evidence artifacts (e.g., policy documents, logs, configuration exports, attestations) remain off-chain in vendor-controlled or mutually agreed secure repositories. On-chain, TPRA records only cryptographic digests (e.g., SHA-256), minimal metadata (type, control mapping, retention tags), content pointers (URI/location reference), and timestamps, enabling later integrity checking and audit reconstruction without storing sensitive contents. The corresponding integrity binding and non-repudiation properties are analyzed in Section~\ref{sec:threat}.}

\textcolor{black}{Given an \textbf{EvidenceRecord}, an assessor (or regulator) retrieves the artifact off-chain, recomputes its digest, and compares it to the on-chain hash; assessor judgments are thereby bound to immutable evidence anchors rather than ambiguous document exchange.}

\subsection{Multi-Assessor Voting and Outcome Construction}
\label{sec:arch:consensus}
\textcolor{black}{TPRA treats assessor disagreement as a primary design constraint. Each assessor submits a structured \textbf{AssessorVote} referencing (i) assessment ID, (ii) control-family/category, (iii) numeric and/or categorical judgments, and (iv) rationale metadata. Votes are auditable ledger objects. Chaincode computes an \textbf{AssessmentOutcome} from the set of submitted votes using a deterministic aggregation rule (e.g., weighted averaging by domain and agreement/dispersion-derived confidence), making the outcome reproducible from immutable vote state.}

\textcolor{black}{Finalization (\textit{completeAssessmentReview}) is endorsement-gated (e.g., Regulator+Assessor), ensuring no single party can unilaterally publish an authoritative compliance report~\cite{gupta2024blockchain}. The endorsement-gated finalization guarantee is formalized as Property~P3 in Section~\ref{sec:threat}.}

\subsection{Auditability, Accountability, and Event-Driven Traceability}
\label{sec:arch:audit}
\textcolor{black}{Beyond immutability, TPRA requires accountable assessment actions: it must be possible to reconstruct who invoked what, which organizations endorsed it, and when it became final. Each chaincode transaction emits audit events (e.g., vendor registered, evidence anchored, vote recorded, outcome finalized) to produce a fine-grained trace aligned with the assessment state machine.}

\textcolor{black}{Formal work on accountability in permissioned blockchains emphasizes that accountability can remain meaningful even under weakened assumptions by enabling identification and blame through undeniable evidence. TPRA aligns with this perspective by ensuring governance-relevant actions (publishing directives, finalizing outcomes) are (i) signed, (ii) endorsed by multiple organizations, and (iii) permanently recorded, supporting dispute resolution using ledger evidence.}

\textcolor{black}{TPRA does not claim that blockchain automatically improves compliance; instead, it specifies a system in which maturity-relevant artifacts---evidence anchors, assessor votes, endorsement satisfaction, and longitudinal state transitions---are cryptographically bound into an auditable history suitable for measuring \emph{framework implementation maturity} over time.}

\section{Framework Implementation Maturity Metrics and Scenario-Driven Evaluation}
\label{sec:eval}

We evaluate whether TPRA supports credible measurement of \emph{framework implementation maturity} (in the sense of capability/maturity levels applied to governed assessment processes) in third-party compliance settings. The evaluation is scenario-driven and audit-trail-based: we instantiate the TPRA lifecycle and then analyze the resulting protocol-enforced artifacts using maturity-relevant metrics. Table~\ref{tab:maturity-metrics} summarizes metrics M1--M8 as the protocol-derived dimensions used to assess maturity, while Table~\ref{tab:maturity-levels} defines qualitative maturity levels used to interpret the combined evidence these metrics provide across repeated assessment cycles.

\textcolor{black}{The metrics in this section therefore emphasize protocol and governance properties (e.g., traceability, multi-party validation, and longitudinal state) rather than isolated performance indicators. They are not intended to guarantee control correctness or security effectiveness; instead, they assess whether a system provides the structural and procedural foundations required for credible maturity measurement beyond static compliance snapshots.}

\begin{table}[!t]
\centering
\caption{Framework Implementation Maturity Metrics and Corresponding TPRA Artifacts}
\label{tab:maturity-metrics}
\renewcommand{\arraystretch}{1.12}
\setlength{\tabcolsep}{6pt}
\begin{tabular}{|c|p{3.0cm}|p{4.25cm}|}
\hline
\textbf{ID} & \textbf{Maturity Metric} & \textbf{TPRA Framework Artifact} \\
\hline
M1 &
Assessment Workflow Integrity &
Ledger-enforced assessment lifecycle with explicit state transitions from assessment initialization to governed finalization. \\
\hline
M2 &
Operationalized Framework Controls &
Canonical decomposition of cybersecurity frameworks into structured control families, assessment questions, and expected evidence types. \\
\hline
M3 &
Identity and Role Enforcement &
Permissioned identity model enforcing role separation among vendors, assessors, and governance authorities. \\
\hline
M4 &
Evidence Traceability &
Cryptographic binding of submitted evidence to specific assessment instances and framework versions. \\
\hline
M5 &
Multi-Assessor Governance &
Preservation of independent assessor evaluations with governed aggregation and consensus-based finalization rules. \\
\hline
M6 &
Longitudinal Trace Persistence &
Persistent assessment records enabling comparison across repeated evaluation cycles and temporal reasoning about control enforcement. \\
\hline
M7 &
Auditability and Transparency &
Immutable recording of assessment actions, evidence references, and finalized compliance outcomes. \\
\hline
M8 &
Structural Cross-Assessment Consistency &
Reuse of canonical assessment structures across vendors and over time without ad hoc redesign or assessor-specific reconciliation. \\
\hline
\end{tabular}
\end{table}

\begin{table}[!t]
\centering
\caption{Framework Implementation Maturity Levels (These levels are ordered and cumulative, like standard maturity models) }
\label{tab:maturity-levels}
\renewcommand{\arraystretch}{1.12}
\setlength{\tabcolsep}{6pt}
\begin{tabular}{|c|p{5.2cm}|}
\hline
\textbf{Level} & \textbf{Description} \\
\hline
Level 0 -- Initial &
Compliance activities are ad hoc, document-based, and assessed independently with limited traceability or reuse. \\
\hline
Level 1 -- Repeatable &
Standardized assessment procedures exist, but evidence validation and assessor coordination remain largely manual. \\
\hline
Level 2 -- Verifiable &
Assessment outcomes are supported by verifiable evidence references and auditable assessment records. \\
\hline
Level 3 -- Governed &
Multi-assessor participation is regulated through defined governance rules and structured assessment finalization. \\
\hline
Level4 -- Continuous &
Compliance is assessed iteratively over time, with persistent state enabling longitudinal analysis and maturity tracking. \\
\hline
\end{tabular}
\end{table}

\subsection{Evaluation Methodology}
\label{sec:eval-method}

This evaluation uses a scenario-driven, metric-based methodology to assess whether TPRA enables maturity-oriented reasoning from protocol traces. Rather than relying on deployment-scale performance measurements, we demonstrate how assessment outcomes are generated, governed, and compared over time using the TPRA protocol (Section~\ref{sec:arch}) and then interpreted using metrics M1--M8 (Table~\ref{tab:maturity-metrics}) and maturity levels L0--L4 (Table~\ref{tab:maturity-levels}). In this paper, L0--L4 primarily serve as descriptive levels for interpreting TPRA traces, while also illustrating how similar evidence-grounded maturity scales may be reused across comparable governed assessment systems.

Specifically, we: (i) construct a realistic third-party assessment scenario in a healthcare RPM setting involving a vendor that provides infrastructure or services to a healthcare organization; (ii) instantiate the TPRA lifecycle, including assessment creation, evidence anchoring, independent assessor voting, and endorsement-gated finalization; and (iii) apply Table~\ref{tab:maturity-metrics} to the resulting ledger artifacts and state across assessment cycles to reason about maturity progression. This methodology emphasizes feasibility, governance enforceability, and longitudinal reasoning, while deferring production deployment benchmarking to future work.

\subsection{Worked Example: Multi-Assessor Vendor Assessment}
\label{sec:eval-worked}

To illustrate TPRA’s support for maturity assessment, we consider a worked example in which a single vendor is assessed across two cycles, denoted $T_1$ and $T_2$. The vendor participates in a healthcare RPM ecosystem and operates cloud-based data ingestion and analytics services for a healthcare provider.

We consider two assessment cycles for the same vendor and scope. At $T_1$, an assessor creates an assessment instance specifying scope, evaluation window, and framework-aligned controls; the vendor submits supporting artifacts (e.g., policy documents, configuration exports, operational records) as off-chain evidence that is registered as on-chain evidence references and cryptographically anchored to the assessment instance. Multiple independent assessors evaluate the evidence and submit votes (potentially divergent) based on their domains. TPRA enforces governed finalization: votes and final outcomes are accepted only under role-based authorization and endorsement constraints, and the finalized assessment (including individual assessor inputs and the governed outcome) is persistently recorded on the ledger. At $T_2$, the same protocol is repeated with updated evidence reflecting operational changes, producing a second finalized assessment under identical governance constraints. Because TPRA preserves assessment state across cycles, the protocol enables explicit $T_1$--$T_2$ comparison (evidence completeness, assessor dispersion, governance enforcement, and outcome evolution) to reason about whether controls were sustained, improved, or degraded over time.

Interpreted through Table~\ref{tab:maturity-levels}, $T_1$ typically aligns with \emph{Repeatable} (L1) or early \emph{Verifiable} (L2), depending on evidence traceability and auditability. At $T_2$, improved evidence-to-control linkage and reduced ambiguity can strengthen assessor convergence and longitudinal comparison, reflecting progression toward \emph{Verifiable} (L2) and, when governed reconciliation is consistently enforced, \emph{Governed} (L3). The key differentiator is that this progression is derived from protocol traces (anchored evidence, recorded votes, and endorsement-gated finalization) rather than static reports.

Applying Table~\ref{tab:maturity-metrics} to the two-cycle trace highlights maturity-relevant properties: workflow integrity (M1), evidence traceability (M4), preservation of independent assessor inputs with governed reconciliation (M5), and longitudinal persistence enabling cross-cycle comparison (M6). In traditional point-in-time audits, prior assessments are typically archived as static reports without structured linkage, limiting their utility for repeatable longitudinal reasoning.

\subsection{Qualitative Scalability Considerations}
\label{sec:eval-qual}

Although this work does not present an empirical performance evaluation, we qualitatively assess whether the framework can scale to realistic third-party compliance settings. The assessment workflow scales primarily with the number of vendors and assessment cycles rather than with high-frequency transaction volume: each assessment instance follows a bounded lifecycle (create, evidence registration, vote submission, finalization). Consequently, ledger interactions grow approximately linearly with assessment events, aligning with periodic reassessment environments typical of compliance programs.

Storage growth is moderated by design: evidence content remains off-chain while cryptographic anchors and metadata are recorded on-chain. This enables long-term audit reconstruction and maturity tracking without storing sensitive artifacts in the ledger state. Permissioned operation with configurable endorsement policies further distributes validation responsibility across organizations while preserving explicit governance constraints.


\section{Threat Model and Security Analysis}
\label{sec:threat}

\textcolor{black}{This section defines the TPRA threat model and argues how the protocol supports integrity, accountability, and governance requirements for multi-organization third-party assessments. Because we evaluate a prototype-level implementation, we focus on protocol-level guarantees, tamper-evident evidence binding, non-repudiation of recorded actions, and endorsement-gated finalization, rather than operational hardening (e.g., sustained DoS resilience).}

\subsection{System Model and Notation}
\label{sec:sys-notation}

We develop TPRA as a permissioned-ledger protocol that executes assessments over repeated cycles. 
An assessment instance for vendor $\mathsf{vid}$ with scope $\mathsf{scope}$ in cycle $i$ is denoted $A_i$. 
Evidence artifacts $E$ are stored off-chain; TPRA records an on-chain anchor $h \leftarrow H(\mathsf{bytes}(E))$, where $H$ is a collision-resistant hash function, and binds $h$ to $A_i$. 
Protocol actions are submitted as signed transactions $(m,\sigma)$, where $m$ is the transaction payload (e.g., action type, parameters, and references to $A_i$) and $\sigma$ is a digital signature under an authenticated identity. 
Critical state transitions (e.g., finalization) are commit-valid only if endorsements satisfy the endorsement policy $\pi$ (e.g., Regulator+Assessor). 
The ledger $\mathcal{L}$ provides an append-only history, and the world state $S$ stores the current assessment state and associated artifacts.

\subsection{Threat Model}
\label{sec:adversary}

\paragraph{Context and motivation}
\textcolor{black}{Third-party compliance workflows often rely on questionnaires and distributed evidence exchange (e.g., email, shared drives, and disparate portals), creating recurring risks: stale approvals as controls degrade over time, fragmented audit trails that prevent reliable reconstruction, and weakly validated self-attestations. TPRA targets these risks by binding evidence and decisions to governed, auditable protocol traces.}

\textcolor{black}{We assume a regulated multi-organization setting in which participants are known and authenticated (permissioned membership), but may act adversarially or strategically.}

\paragraph{Adversary classes}
We consider four adversary classes. A \textcolor{black}{\textbf{malicious vendor ($\mathcal{A}_V$)}} may submit incomplete or misleading evidence, attempt to alter, withdraw, or substitute evidence after submission, or game assessment scope boundaries, thereby undermining the integrity and completeness of the assurance record. A \textcolor{black}{\textbf{malicious assessor ($\mathcal{A}_A$)}} may submit biased votes or rationales, selectively interpret evidence, attempt repudiation of prior actions, or withhold participation to delay finalization, threatening decision integrity and, through delay or non‑participation, system availability. A \textcolor{black}{\textbf{colluding coalition ($\mathcal{A}_C$)}} may coordinate to influence outcomes (e.g., coordinated voting), but is bounded by endorsement and quorum thresholds, so its primary effect is on the integrity of aggregated decisions. Finally, an \textcolor{black}{\textbf{off-chain attacker ($\mathcal{A}_O$)}} targets evidence storage to delete, replace, rollback, or deny access to evidence objects without controlling on-chain identities, directly threatening evidence integrity and availability. 

\paragraph{Attack Surfaces}
\label{sec:attack-surfaces}

\textcolor{black}{We do not treat Fabric consensus safety or ledger immutability as primary attack vectors; instead, we focus on practical, in-scope surfaces: chaincode workflow correctness, client/API misuse, endorsement and identity configuration, and off-chain evidence integrity and availability. Sustained DoS is an operational concern and is not evaluated here.} Concretely, the primary attack surfaces include the \textcolor{black}{\textbf{client-to-ledger submission interfaces}}, where adversarial inputs or API misuse may target validation logic and workflow preconditions; the \textcolor{black}{\textbf{chaincode state machine}}, where missing guards or implementation errors could enable unintended transitions (e.g., premature finalization); and the \textcolor{black}{\textbf{endorsement/ACL configuration}}, where mis-scoped or overly permissive policies can weaken multi-party authorization. We also consider \textcolor{black}{\textbf{identity and key management}} risks, such as compromised keys or mis-issued certificates enabling validly signed malicious actions, as well as \textcolor{black}{\textbf{off-chain evidence repositories}}, which may be targeted for deletion, replacement, rollback, or denial of access. Finally, we include \textcolor{black}{\textbf{evidence reference binding}} attacks (e.g., Uniform Resource Identifier (URI) swapping or stale references), mitigated by binding anchors to $A_i$ and recording minimal metadata, and \textcolor{black}{\textbf{governance/finalization operations}} where an adversary attempts to bypass role checks or endorsement-gated finalization.

\subsection{Assumptions, Scope, and Security Objectives}
\label{sec:assumptions-objectives}

\paragraph{Assumptions}
Our analysis relies on standard cryptographic and platform assumptions: \textcolor{black}{\textbf{A1 (Hash security):}} $H(\cdot)$ is collision resistant; \textcolor{black}{\textbf{A2 (Signature security):}} the signature scheme is EUF-CMA secure; and \textcolor{black}{\textbf{A3 (Endorsement correctness):}} a transaction is committed only if endorsements satisfy $\pi$ and all signatures verify. We further assume the permissioned platform correctly enforces membership, endorsement validation, and append-only history.

\paragraph{Scope and out-of-scope}
Physical-device attacks and side channels are out of scope. TPRA does not determine the factual truthfulness of evidence contents; rather, it provides tamper-evidence, attribution, and auditability for what was submitted and how decisions were reached. Availability is treated as an operational concern: multi-organization replication supports access to history, while sustained DoS is not evaluated.

\paragraph{Security objectives and mechanisms}
TPRA targets protocol-level properties required for credible third-party assessment under governance. \textcolor{black}{\textbf{O1--O3 (Confidentiality, integrity, traceability):}} evidence contents remain off-chain under repository/IAM controls, while the ledger stores anchors and minimal metadata and chaincode constrains lifecycle transitions, yielding tamper-evident assessment state and evidence provenance. \textcolor{black}{\textbf{O4 (Accountability/non-repudiation):}} all actions are identity-bound signed transactions recorded on an append-only ledger. \textcolor{black}{\textbf{O5 (Governed finalization):}} endorsement/quorum policies enforce multi-party authorization for critical actions (e.g., publishing directives and finalizing outcomes), preventing unilateral commits. \textcolor{black}{\textbf{O6 (Availability, operational):}} multi-organization replication improves resilience and access to history, but sustained DoS is not evaluated.

\subsection{Formal Security Properties and Proof Sketches}
\label{sec:formal-security}

\textcolor{black}{We provide compact, game-based sketches for three core protocol properties.}

\paragraph{\textcolor{black}{P1: Evidence binding (tamper-evident anchoring)}}
\textcolor{black}{TPRA provides evidence binding if, after an anchor $h$ is committed for an assessment instance $A$, no adversary can later present a different evidence object as the one anchored by $h$ without detection.}

\textcolor{black}{\textbf{Game $\mathsf{G}_{\textsf{Bind}}$ (Evidence Substitution).}} \textcolor{black}{The adversary selects an assessment identifier $A$ and submits evidence $E$. The protocol computes $h \leftarrow H(\mathsf{bytes}(E))$ and commits $(A,h)$. The adversary wins if it outputs $E' \neq E$ such that $H(\mathsf{bytes}(E')) = h$.} \textcolor{black}{\textbf{Proof sketch.}} \textcolor{black}{If the adversary wins $\mathsf{G}_{\textsf{Bind}}$, then it has produced $E' \neq E$ such that $H(\mathsf{bytes}(E')) = H(\mathsf{bytes}(E))$, which constitutes a hash collision (and more specifically a second-preimage) for $H(\cdot)$. Under Assumption~\textbf{A1}, the adversary's success probability is negligible.}



\paragraph{\textcolor{black}{P2: Non-repudiation of assessment actions}}
\textcolor{black}{TPRA provides non-repudiation if committed actions cannot be falsely attributed to an honest participant who did not authorize them, and an authorizing participant cannot later deny having signed the action.}

\textcolor{black}{\textbf{Game $\mathsf{G}_{\textsf{NR}}$ (Action Forgery).}}
\textcolor{black}{The adversary outputs a committed record $(m,\sigma,pk)$ such that $\mathsf{Verify}_{pk}(m,\sigma)=1$ for an honest public key $pk$, and $m$ was not signed by the corresponding participant.}

\textcolor{black}{\textbf{Proof sketch.}}
\textcolor{black}{Winning $\mathsf{G}_{\textsf{NR}}$ yields a valid signature on a message not signed by the honest key owner, which contradicts EUF-CMA security (\textbf{A2}); thus the adversary's success probability is negligible.}




\paragraph{\textcolor{black}{P3: Governed finalization (endorsement-gated outcomes)}}
\textcolor{black}{TPRA provides governed finalization if a final outcome cannot be committed unless the endorsement policy $\pi$ is satisfied.}

\textcolor{black}{\textbf{Game $\mathsf{G}_{\textsf{Gov}}$ (Policy Bypass).}}
\textcolor{black}{The adversary controls a set of identities whose endorsements do not satisfy $\pi$ (i.e., it lacks the required endorsing organizations) and wins if it causes a \textsf{FinalizeOutcome} transaction to be committed without the endorsements required by $\pi$.}

\textcolor{black}{\textbf{Proof sketch.}}
\textcolor{black}{Under \textbf{A3}, a transaction is committed only if the collected endorsements satisfy $\pi$ and all signatures verify; hence any successful bypass would require either (i) violating endorsement-policy enforcement (breaking \textbf{A3}) or (ii) forging endorsements via signature forgery (contradicting \textbf{A2}). Therefore, the adversary's success probability is negligible.}

\subsection{Adversarial Considerations and Limitations}
\label{sec:limitations}

\textcolor{black}{TPRA strengthens integrity, accountability, and governed outcome construction, but it does not eliminate all adversarial risks. Collusion can still influence outcomes if governance thresholds are misconfigured or if sufficient endorsers collude. TPRA also cannot independently verify the factual correctness of off-chain evidence; it provides tamper-evidence, attribution, and auditability for what was submitted and how it was decided. Privacy depends on secure off-chain repositories and access controls, which are assumed but not evaluated here. These limitations motivate future work on stronger collusion resistance, privacy-preserving evidence validation (e.g., selective disclosure), and formal verification of chaincode workflow correctness.}

\section{Discussion and Future Work}

\textcolor{black}{This section summarizes the evaluation implications of TPRA, states scope limitations of our study design, and outlines high-impact next steps.}

\paragraph{Comparative Discussion}\textcolor{black}{Traditional third-party audits are largely point-in-time: evidence is summarized in static reports and prior assessments are not structurally linked to future evaluations. Centralized GRC platforms improve workflow, but typically concentrate interpretation and outcome construction within a single authority, limiting multi-party governance and independently constructed outcomes across organizational boundaries. TPRA reframes assessment as a \emph{system property} by enforcing assessment semantics through chaincode, requiring endorsement-gated finalization, and preserving assessment state across cycles. The contribution is not new controls or scoring criteria, but a protocol substrate that produces protocol-enforced artifacts (lifecycle state, evidence anchors, assessor inputs, and governed finalization) from which maturity can be inferred, reproduced, and compared longitudinally. The proposed metrics and maturity levels provide a compact basis for comparing third-party compliance systems on governance enforceability, reproducibility, and cross-cycle traceability.}

\paragraph{Future Work}\textcolor{black}{Next steps include applying the methodology to additional regulated domains. In companion work, \emph{TPRA-Trust}, we will extend TPRA with a protocol-derived trust model for assessor disagreement. The model will trust-weight assessor influence during outcome construction, while remaining bounded by quorum/endorsement and appeal constraints. It will also derive longitudinal trust-stability signals across repeated assessment cycles as an additional maturity indicator. Trust values will be computed from ledger-observable behavior (e.g., participation reliability, consistency with prior validated outcomes, and dispute/appeal history). These values will be stored as auditable state reproducible from audit trails. We defer full trust-model definition, parameterization, and robustness analysis (including collusion and strategic behavior) to the TPRA-Trust paper.}

\section{Conclusion}

This paper introduced a maturity-oriented evaluation approach for blockchain-based third-party compliance assessment systems. Rather than treating compliance as a point-in-time outcome, we frame third-party assessment as a longitudinal measurement problem that requires verifiable evidence handling, governed validation, and persistent assessment state across organizational boundaries.

We defined framework implementation maturity metrics and a qualitative maturity model grounded in protocol-enforced artifacts (anchored evidence references, preserved assessor inputs, endorsement-gated finalization, and immutable audit traces). Using TPRA as an instantiation in a healthcare remote patient monitoring setting, we demonstrated how these constructs support reproducible and auditable reasoning across assessment cycles. 

This work does not argue that blockchain universally improves compliance; it identifies when a permissioned ledger is most valuable, when outcomes must be independently validated, tamper-evident, and governed across multiple organizations over time. The proposed metrics and maturity model are domain-agnostic and provide a foundation for designing and comparing third-party compliance systems using maturity-oriented criteria.

{\small
\bibliographystyle{IEEEtran}
\bibliography{references}
}

\end{document}